\documentclass[12pt]{article}
\usepackage{psfig,epsfig}
\usepackage{latexsym}
\usepackage{pifont}

\newcommand{\lapprox}{\lower.6ex\hbox{$\; \buildrel<\over\sim \;$}}

\newcommand{\gapprox}{\lower.6ex\hbox{$\; \buildrel>\over\sim \;$}}

\newcommand{\curly}{\lower1.ex\hbox{$\; \stackrel{\textstyle \wr}{\wr} \;$}}

\def\barr{\begin{array}}
\def\earr{\end{array}}
\def\berr{\begin{eqnarray}}
\def\err{\end{eqnarray}}
\def\berrno{\begin{eqnarray*}}
\def\errno{\end{eqnarray*}}
\def\be{\begin{equation}}
\def\ee{\end{equation}}

\def\fr{\frac}

\def\apj{{\it Astrophys.~J.}}
\def\aj{{\it Astronom.~J.}}
\def\apjl{{\it Astrophys.~J.~Lett.}}
\def\prd{{\it Phys.~Rev.~D.}}
\def\prl{{\it Phys.~Rev.~Lett.}}
\def\mnras{{\it Mon. Not. R.~Astr.~Soc.}}
\def\nature{{\it Nature}}

\def\AnA{{\it Astr. Astrophys.}}
\def\grg{{\it Gen. Rel. Grav.}}

\renewcommand{\a}{\alpha}

\newcommand{\g}{\gamma}
\newcommand{\G}{\Gamma}

\renewcommand{\L}{\Lambda}

\renewcommand{\O}{\Omega}

\renewcommand{\t}{\theta}
\renewcommand{\v}{\varphi}

\title{Cosmological Tests for a Linear Coasting Cosmology}
\author{Abha Dev\thanks{E--mail : abha@ducos.ernet.in},
Margarita Safonova\thanks{E--mail : rita@ducos.ernet.in}, 
Deepak Jain\thanks{E--mail : deepak@ducos.ernet.in}, and
Daksh Lohiya\thanks{E--mail : dlohiya@ducos.ernet.in} \\
       {\em Department of Physics and Astrophysics,} \\
       {\em University of Delhi, Delhi-110007, India}
       }
\begin{document}

\maketitle
 
\begin{abstract}

A strictly linear evolution of the scale factor
is a characteristic feature in several classes of alternative gravity 
theories. In this article we investigate the overall viability of an open
linear coasting cosmological model. We 
report that this model is consistent with gravitational lensing 
statistics (within $1\sigma$) and accomodates old high-redshift galaxies. 
We finally conclude that such a linear coasting, $\a(t) = t$, is not ruled out
on basis of these observational tests.
 
\end{abstract} 

\section{Introduction}

%% Problems in SM.

Standard cold dark matter (CDM) cosmology
presents serious theoretical and observational problems as a model
for an acceptable description of the Universe. This has motivated a search for
 alternative cosmological models \cite{alternative}. The first problem comes
from the conflict between age of the Universe and the age of the oldest stars
in Galactic globular clusters. The age constraints from old galaxies at
high redshifts render ``the age problem" even more acute \cite{Krauss}. 

Another major  difficulty is the cosmological constant ($\L$) problem. While it is 
difficult to have a decent theoretical justification for $\L$ 
\cite{models}, there
are serious doubts on the compatibility of the Hubble Deep Field (HDF) data 
with large $\L$. Initial analysis of Maoz and Rix \cite{Maoz1}
placed a somewhat stringent upper limit of $\Omega_{\L} < 0.7$ 
which have been somewhat softened by later analysis of Cooray et al. \cite{Cooray}.

These problems have generated a lot of interest in an open FRW model with a linear
evolution of the scale factor, $a(t) \propto t$. In such a cosmology the
universe expands with constant velocity; hence the term coasting cosmology
\cite{Kolb}. Notable among such models is a recent idea of Allen \cite{allen},
in which  such a scaling results in an $SU(2)$ cosmological instanton dominated
universe. The Weyl gravity theory of Manheim and Kazanas \cite{mann} makes space 
for yet another possibility. Here again the FRW scale factor approaches a
linear evolution at late times.   

%% Motivations for us
 
The need for investigating such a model comes from 
several considerations. Particle horizons occur in models with $a(t) \approx t^{\a}$ for $\a
<1$ and, therefore, linear coasting model does not suffer the horizon
problem. Also, linear evolution of a scale factor is supported in alternative
gravity theories (eg. non-minimally coupled scalar-tensor theories), where it
turns out to be independent of the matter equation of state (see \cite{meetu} and
references therein). The scale factor
in such theories does not constrain the matter density parameter and,
therefore, does not present any flatness problem. Moreover, the age of the 
coasting universe is 1.5 times the age of standard CDM
universe \cite{meetu}, which makes this model comfortably concordant with 
the ages of globular clusters. Finally, a linear
coasting cosmology, independent of the equation of state of matter, is a
generic feature in a class of models that attempt to dynamically solve the
$\L$ problem \cite{models}. Such models have a scalar field non-minimally
coupled to the curvature of the universe. With the evolution of time, the
non-minimal coupling diverges, the scale factor quickly approaches linearity
and the non-minimally coupled field acquires a stress energy that cancels the
vacuum energy in the theory.

Interestingly it was noted by Perlmutter et al. \cite{perl} that the curve for 
$\O_{\rm M} = \O_{\rm \L}
= 0$ (for which the scale factor would have linear evolution) is ``practically 
identical to the best fit plot for an unconstrained cosmology''. Recently, it was shown by 
Dev et al. \cite{abha} that open linear coasting cosmology presents a good fit to
the SNe Ia data. It was also demonstrated that this model is consistent 
with the primordial nucleosynthesis \cite{annu}.    

In this paper we consider constraints on the index $\a$ of a power law cosmology,  
$a(t) \propto t^{\a}$, from two different tests: gravitational 
lensing statistics and age 
estimates of old high-redshift galaxies. The expected frequency of multiple
imaging lensing events is a sensitive probe for the viability of a given
cosmology. In view of the successful results of the above mentioned works
\cite{abha,annu}, we used this test to constrain the power index $\alpha$ of
the scale factor. Expected number of lens systems depends upon the index
$\alpha$ through the angular diameter distances. By varying $\alpha$, the
number of lenses changes which on comparison with the observations gives us
the constrain on $\alpha$. Age measurements of old high-redshift galaxies
give lower bound on the power index $\alpha$. This is based on the fact that
the age of the universe in a given redshift is greater than or at least equal
to the age of its objects. Since the age of the universe is a function of
$\alpha$, we find the value of $\a$ which permits the existence of these old
galaxies. 

In Section 2 we introduce  the ansatz for the power law cosmology and derive
the angular diameter distance formula. In Section 3 we describe the tests
and the constraints they present on the power index for that cosmology. We
summarize our results and present the result of a joint constraint from various
observational tests in Section 4. 

\section{Linear Coasting Cosmology}
We consider a general power law cosmology with the scale factor given in
terms of two arbitrary dimensionless parameters $B$ and $\a$
\be
a(t)=B\fr{c}{H_0}\left(\fr{t}{t_0}\right)^{\a}\,\,,
\label{eq:ansatz}
\end{equation}
for an open FRW metric  
\be
ds^2=c^2dt^2-a^2(t) \left[\frac{dr^2}{1 + r^2} + r^2(d\t^2
+\sin^2{\t}\,d\phi^2)
\right]\,\,.
\end{equation}
Here $t$ is cosmic proper time and $r,\,\t,\,\v$ are comoving spherical 
coordinates. 

The expansion rate of the universe is described by a Hubble parameter, $H(t) =
\dot{a}/a ={\a}/{t}$. The {\it present} expansion rate of the universe is
defined by a Hubble {\it constant}, equal in our model to $H_0=\a/t_0$ (here
and subsequently the subscript 0 on a parameter refers to its present value).
The scale factor and the redshift are related to their present values by
$a/a_0=(t/t_0)^{\a}$.  As usual, the ratio of the scale factor at the emission
and absorption of a null ray determines the cosmological redshift $z$ by
\be
\frac{a_0}{a(z)} = 1+z\,\,,
\label{eq:redshift}
\end{equation}
and the age of the universe is
\be
t(z)=\fr{\a}{H_0(1+z)^{1/\a}}\,\,.
\label{eq:age}
\end{equation}
Using (\ref{eq:redshift}), we define the dimensionless Hubble parameter   
\be
h(z) \equiv \frac{H(z)}{H_0}=(1+z)^{1/\a}.
\label{eq:dimensionless_hubble}
\end{equation}
The present `radius' of the universe is defined as (see Eq.~\ref{eq:ansatz}) 
\be
a_0 =B\fr{c}{H_0}\,\,.
\label{eq:a_0}
\end{equation}

In terms of the parameters $\a$ and $B$, the angular diameter distance
between two different redshifts is: 
\be
D_{\rm A}(z_1,z_2 ,\a) = \fr{Bc}{(1+z_2) H_0}
\sinh{\left[ \fr{1}{B}\frac{\a}{\a-1}
\left\{ (1+z_2)^{\frac{\a-1}{\a}} - (1+z_1)^{\frac{\a-1}{\a}} 
\right\} \right]}\,\,.
\label{DA}
\end{equation} 
In a limiting case,  $\a \rightarrow 1$, we obtain the following expression
\be
D_{\rm A}(z_1,z_2) = \fr{Bc}{2 H_0}\fr{\left[(1+z_2)^{2/B} -
(1+z_1)^{2/B} \right]}{(1+z_1)^{1/B}(1+z_2)^{\fr{B+1}{B}}}\,\,.
\end{equation}

The look-back time, which is the difference between the age of the universe
when a particular light ray was emitted and the age of the universe now, we
find as  
\be
\fr{c\,dt}{dz_{\rm L}}=\fr{c}{H_0 (1+z_{\rm L})^{\fr{\a
+1}{\a}}}\,\,. 
\label{eq:backtime}
\end{equation}    

\section{Testing the model against observations}

\subsection{Gravitational lensing statistics}

We consider a sample of 867 ($z > 1$) high luminosity
optical quasars which include 5 lensed quasars ($1208 +1011$, 
H $1413+117$, LBQS $1009+0252$, PG $1115+080$, $0142+100$). This sample 
is taken from optical lens surveys such as the HST Snapshot survey \cite{HST}, 
the Crampton survey \cite{Crampton}, the Yee survey \cite{Yee}, Surdej
survey \cite{Surdej}, the NOT Survey \cite{Jaunsen} and the FKS survey
\cite{FKS}. The lens surveys and quasar catalogs usually use V magnitudes, so
we transform $m_{V}$ to a B-band magnitude using an average $B-V$ colour of
$0.2$ mag \cite{Bahcall,Maoz,Maoz2}.

The differential probability $d\tau$ of a beam having a lensing event in
traversing $dz_{\rm L}$ is
\be
d\tau =  n_0(1 + z_L)^{3}\,\sigma \,\frac{cdt}{dz_{\rm L}} dz_{\rm L}\,,
\label{dtau1}
\end{equation}
where $n_0$ is the present comoving number density of the lenses, $\sigma$ 
is the cross-section for lensing event and the quantity $cdt/dz_{\rm L}$ 
is given by (\ref{eq:backtime}).

For simplicity we use the Singular Isothermal Sphere (SIS) model for the 
lens mass 
distribution. The cross-section for lensing events for the SIS model is given
by \cite{TOG}
\be
\sigma = 16\,\pi^{3}\,\left({v \over c}\right)^{4}\left(\frac{D_{\rm OL}
D_{\rm LS}}{D_{\rm OS}}\right)^{2}\,, \label{sigma}
\ee
where $v$ is the velocity dispersion of the dark halo of the lensing
galaxy. We define $D_{\rm
OL},\,D_{\rm OS}$ and $D_{\rm LS}$ as the angular diameter distances from the
observer to the lens, to the source and between the lens and the source,
respectively. 

Assuming no evolution of the galaxies, the comoving number density is modeled
by Schechter function as \begin{equation}
\Phi(L,z = 0)\,dL = \phi_\ast \,\,\left(\fr{L}{L_\ast}\right)^{\tilde \a}
\,\exp{\left(-\fr{L}{L_{\ast}}\right)}\,\fr{dL}{L_{\ast}}\,\,,
\label{ro4}
\end{equation}

where $\phi_{\ast}$, $\tilde \a$ and $L_{\ast}$ are the normalization factor, the
index of faint-end slope, and the characteristic luminosity, respectively.
We also assume that the velocity dispersion of dark matter halo $v$ is
related to the luminosity $L$ by the Faber-Jackson relation for E/S0 galaxies
$v=v_{*} \left(L/L_{\ast}\right)^{1/\gamma}$.

The differential probability is given by\cite{ffkt}
\be
d\tau = F^{*}(1 + z_{\rm L})^{3}\fr{H_0}{c}\left(\fr{H_0D_{\rm OL}
D_{\rm LS}}{c\,D_{\rm OS}}\right)^{2}{cdt \over dz_{\rm L}} dz_{\rm L}\,\,,
\ee
where
\be
F^* = {16\pi^{3}\over{c
H_{0}^{3}}}\phi_\ast v_\ast^{4}\Gamma\left(\tilde \alpha + {4\over\gamma} +1\right)\,
\ee
is the dimensionless quantity, which measures the effectiveness of matter in
producing multiple images. Table~1 lists Schechter and lens parameters for E/S0
galaxies, as suggested by \cite{loveday} (hereafter $LPEM$ parameters).
We neglect the contribution of spirals as lenses, as their velocity dispersion 
is small in comparison to E/S0 galaxies.

The constraints obtained on the cosmological paramters are highly
dependent on the choice of lens and Schechter paramters. 
The lens and Schechter parameters should be determined in a highly
correlated manner from a
galaxy  survey. The use of parameters derived from various surveys
might introduce error. We consider LPEM parameters as they
form one such set of paramters and they also take into
account the morphological distribution of the E/S0 galaxies\cite{c}.
Recently several galaxy surveys have come up with a much larger
sample of galaxies. This has improved our knowledge of the galaxy
luminosity function. However, these surveys
donot classify the galaxies by their morphological type \cite{lc,blanton,cross}.

The differential optical depth of lensing in traversing $dz_{\rm L}$ with
angular separation between $\phi$ and $\phi + d\phi$ reads as 
\berr
%\hspace{-1 cm} 
\lefteqn{\fr{d^2\tau}{dz_{\rm L}d\phi}d\phi dz_{\rm L}=}\nonumber \\
&&\hspace{-0.7 cm}F^{*}(1+z_{\rm L})^3\fr{H_0}{c}
\fr{\g/2}{\G \left(\tilde \a +1+ \fr{4}{\g}\right)}\fr{cdt}{dz_{\rm L}}
\left(\fr{H_0 D_{\rm L}D_{\rm LS}}{c\,\, D_{\rm S}}\right)^2
\left(\fr{D_{\rm S}}{D_{\rm LS}}\phi\right)^{\fr{\g}{2}(\tilde \a + 1 +\fr{4}{\g})}
\exp{\left[\left(-\fr{D_{\rm S}}{D_{\rm LS}}\phi \right)^{\fr{\g}{2}}\right]}
\fr{d\phi}{\phi}dz_{\rm L}\,,\nonumber \\
&& 
\label{eq:diff}
\err
where $\phi = \Delta\theta/8\pi (v_{\ast}/c)^{2}$ and $v_{\ast}$ the velocity
dispersion corresponding to the characteristic luminosity $L_{\ast}$ in
(\ref{ro4}).

We make two corrections to the optical depth to get the lensing
probability: magnification bias and selection function. Magnification
bias, ${\bf B}(m,z)$, is to take into account the increase in the apparent
brightness of a quasar due to lensing, which, in turn, increases the
expected number of lenses in flux limited sample.

The bias factor for a quasar at redshift $z$ with apparent magnitude
$m$ is given by \cite{Turner,1CSK,2CSK}
\be
{\bf B}(m,z) = M_{0}^{2}\, \emph{B}(m,z,M_{0},M_{2})\,\,,
\label{bias}
\ee
where
\be
\emph{B}(m,z,M_{1},M_{2})= 2\,\left(\frac{dN_{\rm Q}}{dm}\right)^{-1}
\int_{M_{1}}^{M_{2}}\frac{dM}{M^{3}}
\frac{dN_{\rm Q}}{dm}\left(m+2.5\log(M),z\right).
\label{bias1}
\end{equation}
In the above equation $\left({dN_{\rm Q}(m,z)}/{dm}\right)$ is the measure of
number of quasars with magnitudes in the interval $(m,m+dm)$ at redshift $z$.
We can allow the upper magnification cutoff $M_{2}$ to be infinite, though in
practice we set it to be $M_{2} = 10^{4}$. $M_{0}$ is the minimum magnification
of a multiply imaged source and for the SIS model $M_{0} =2$.

We use Kochanek's ``best model'' (K96) for the quasar luminosity
function:
\be
\frac{dN_{\rm Q}}{dm}(m,z) \propto
\left(10^{-a(m-\overline{m})}+10^{-b(m-\overline{m})}\right)^{-1}\,\,,
\label{lum}
\ee
where the bright-end slope $a$ and faint-end slope $b$ are constants, and the
break magnitude $\overline{m}$ evolves with redshift:
\be
\overline{m} = \left\{ \begin{array}{ll}
                   m_{o}+(z-1)    & \mbox{for $z < 1$}, \\
                   m_{o}          & \mbox{for $1 < z \leq 3$}, \\
                   m_{o}-0.7(z-3) & \mbox{for $z > 3$}.
                   \end{array}
               \right. \
\ee
Fitting this model to the quasar luminosity function data in \cite{Qdata} for
$z > 1$, Kochanek finds that ``the best model'' has $a = 1.07 \pm 0.07$, $b =
0.27 \pm 0.07$  and $m_{o} = 18.92 \pm 0.16$ at B magnitude. The magnitude
corrected probability, $p_{i}$, for the quasar $i$ with apparent magnitude
$m_{i}$ and redshift $z_{i}$ to get lensed is: 
\be
p_{i} = \tau(z_{i}){\bf B}(m_{i},z_{i})\,.
\label{prob1}
\ee

Selection effects are caused by limitations on dynamic range, limitations on
resolution and presence of confusing sources such as stars. The survey can
only detect lenses with magnifications larger than $M_{f}$. This sets the
lower limit on the magnification. Therefore the $M_{1}$ in the bias function
(\ref{bias1}) gets replaced by $M_{f}(\theta)$ (for details, see K93),
which is given as \be
M_{f} = M_{0}(f+1)/(f-1)\,\,,
\ee
with
\be
f = 10^{0.4\,\Delta m(\theta)}\,\,.
\ee

The corrected lensing probability and image separation distribution function
for a single source at redshift $z_{\rm S}$ are given in K96
\be
p^{'}_{i}(m,z) = p_{i}\int \frac{ d(\Delta\theta)\,
p_{c}(\Delta\theta)\emph{B}(m,z,M_{f}(\Delta\theta),M_{2})}
{\emph{B}(m,z,M_{0},M_{2})},
\label{prob2}
\ee
and
\be
p^{'}_{ci} =p_{ci}(\Delta\theta)\,\frac{p_{i}}{p_{i}^{'}}\,
\frac{\emph{B}(m,z,M_{f}(\Delta\theta),M_{2})}
{\emph{B}(m,z,M_{0},M_{2})}\,\,,
\label{confi}
\ee
where
\be
p_{c}(\Delta\theta) =
\frac{1}{\tau(z_{S})}\,\int_{0}^{z_{S}}\frac{d^{2}\tau}{dz_{L}d(\Delta\theta)}
\,dz_{L}\,\,.
\label{pcphi}
\ee
Equation (\ref{confi}) defines the configuration probability. It is the
probability that the lensed quasar $i$ is lensed with the observed image
separation.

To get selection function corrected probabilities, we divide our sample into
two parts---the ground based surveys and the HST survey. We use the selection
functions as suggested in K93.

In our present calculations we do not consider the extinction effects due to
the presence of dust in the lensing galaxies.

\begin{figure}[ht]
\vspace{-1.2in}
\centerline{
\epsfig{figure=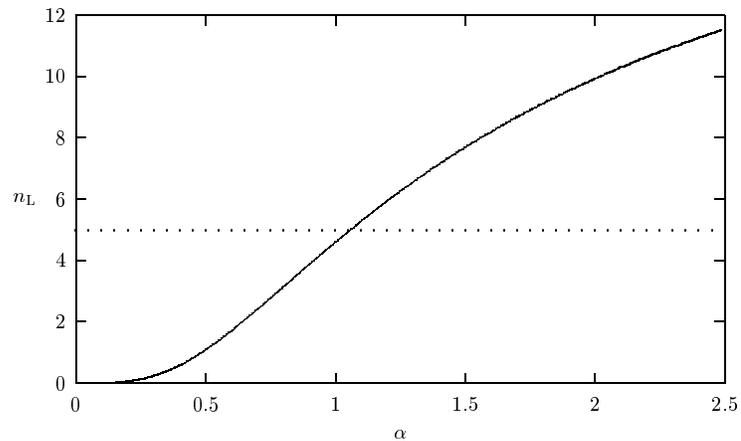,width=1.\textwidth}}
\vspace{-5.2in} 
\caption{Predicted number of lensed quasars $n_{\rm L}$ in the adopted optical quasar sample,
 with image separation $\Delta\theta \leq 4$, Vs power index $\a$.}
\label{fig:NlA}
\end{figure}

Finally, we use the above basic equations to perform the following tests :

\begin{trivlist}
\item(i) The sum of the lensing probabilities $p_{i}^{'}$ for the optical 
QSOs gives the
expected number of lensed quasars, $n_{\rm L} = \sum\, p_{i}^{'}$. The
summation is over the given quasar sample.We look for those values of
the parameter for which the adopted optical sample has exactly five lensed
quasars (that is those values of the parameters for which $n_{\rm L} =
5$). 

We start with a two parameter fit. We allow $\a$ to vary in the range ($0.0 \leq \a
\leq 2.0$) and $B$ to vary in the range ($0.5\leq B \leq 10.0$). We observe that 
for $B\ge 1$, $D_A$ becomes independent of it. It can be easily checked from
equation (\ref{DA}) that for large values of $B$, $\sinh(x) \sim x$ making
$D_A$ independent of $B$. From the previous works constraining power law cosmology
\cite{abha,annu}, the value of $B = 1$ is found to be compatible with
observations. Incidentally, we can estimate the present scale factor of  the
universe as $a_0 \approx c/H_0$, hence we use $ B = 1$ in further analysis.  

Fig.~\ref{fig:NlA} shows the predicted number of lensed quasars for the above 
specified range of $\alpha$. We obtain $n_{\rm L} = 5$ for $\a = 1.06$. We
further  generate $10^{4}$ quasar samples (each sample has 867 quasars)
using bootstrap method and find best fit
$\a$ for  each data set to obtain error bars on $\a$. We finally obtain $\a =
1.09 \pm 0.3$.

\item(ii) We also perform maximum likelihood analysis to determine the value 
of $\alpha$, for which the observed sample becomes the most probable observation. 
The likelihood function is
\be
{\cal{L}} = \prod_{i=1}^{N_{\rm U}}(1-p^{'}_{i})\,\prod_{k=1}^{N_{\rm L}}
p_{k}^{'}\,p_{ck}^{'}\,\,.
\label{LLF}
\ee
Here $N_{\rm L}$ is the number of multiple-imaged lensed quasars, $N_{\rm U}$
is the number of unlensed quasars, $p_{k}^{'}$, the probability of quasar $k$
to get lensed is given by Eq.~\ref{prob2} and $p_{ck}^{i}$, the configuration
probability, is given by Eq.~\ref{confi}. The best fit (${\mathcal{L}}_{\rm
max}$) occurs for $\a = 1.13$. We see that $0.85 \leq \a \leq 1.56$ at
$1\sigma$ ($68\%$ confidence level) and $0.65 \leq \a \leq 2.33$ at $2\sigma$
($95.4\%$ confidence level).
\end{trivlist}

\noindent However, gravitational lensing statistics is susceptible to
a number of uncertainties. The constraints obtained on the
cosmological parameters from the statistics of strong lensing may vary
after inclusion of the uncertainties in the luminosity function,
lensing cross-section for galaxies (E/S0) and  quasar luminosity
function, role of spirals and the dust extinction.

\subsection{Constraints from age estimates of high-z galaxies}

Another observational test which can constraint $\alpha$ is the
age measurement of the old high redshift galaxies (OHRG)\cite{Lima,Lima2}.
These constraints are more stringent than those obtained from globular
cluster age measurements \cite{Dunlop,Krauss,RH}. Here, we
consider the galaxy 3C65 at $z=1.175$ (4 Gyr old) \cite{Stockton}, 
at $z=1.55$ (3.5 Gyr old; 53W091) \cite{Dunlop,Spinrard} and a 4 Gyr old galaxy 
53W069 at $z=1.43$ \cite{Dunlop2}. These are the minimum ages of these galaxies as
indicated by best fitting spectral synthesis models. The age of the universe at a 
given redshift is greater than or at least equal to the age of its oldest objects 
at that redshift. In power law cosmologies the age of the universe increases 
with increasing $\alpha$. Hence this test gives lower bound on $\alpha$. This 
can be checked if we define the dimensionless ratio:
\be
{t(z)\over t_{\rm g}} = {f(\alpha, z)\over H_0 t_{\rm g}} \ge 1\,\,.
\ee
The $t_{\rm g}$ is the age of an old object and $f(\alpha, z)=\a/(1+z)^{1/\a}$.
The error bar on $H_{0}$ determines the extreme value of $t_{\rm g}$. The lower
limit on $H_0$ was recently updated to nearly $10\%$ of accuracy by Freedman
\cite{Freedman}: $H_0 = 70 \pm 7$ km/sec/Mpc; $1\sigma$. The constraints from 
SNe Ia data also point to $H_0 > 60$ km/sec/Mpc \cite{Riess}. For the galaxy 3C65, 
the lower limit on age (4.0 Gyr) yields: $0.26 \leq H_0\,t_{\rm g} \leq 0.32$. 
Similarly for the galaxy 53W069 we have $0.26 \leq H_0\,t_{\rm g} \leq 0.32$. 
For the galaxy 53W091 at $z = 1.55$ the lower limit on age (3.5 Gyr) gives 
$0.23 \leq H_0\,t_{\rm g} \leq 0.28$.   

\begin{figure}[ht]
\centerline{
\epsfig{figure=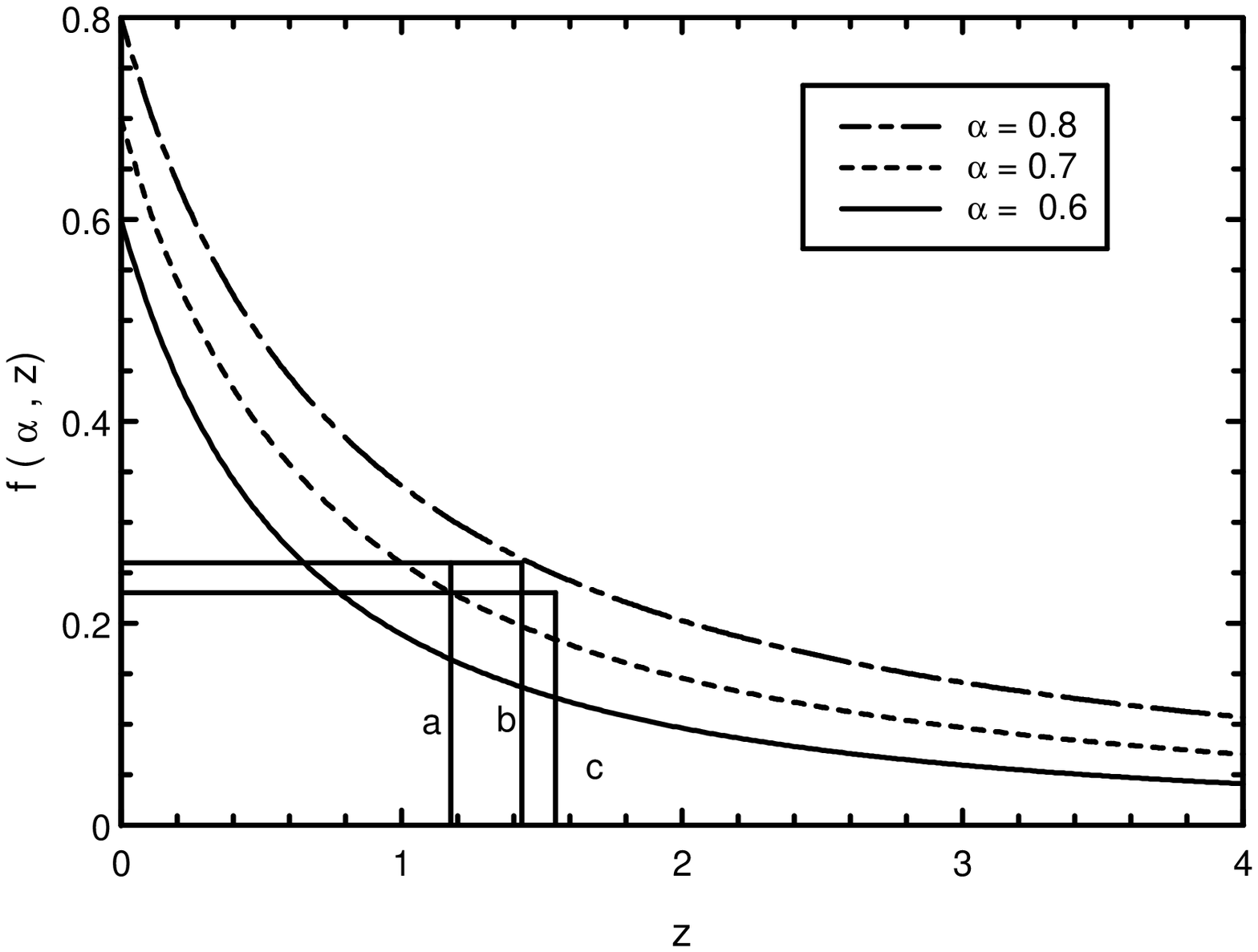,width=.6\textwidth}}
\vspace{-1.2in} 
\caption{$f(\a,z)$ vs. $z$ for various different values of $\a$; 'a'
corresponds 3C65 galaxy ($z = 1.175$), 'b' corresponds to the galaxy
53W069 ($z = 1.43$) and 'c' corresponds to 53W091 ($z = 1.55$).} 
\label{fig:OHRG}
\end{figure}

Fig.~\ref{fig:OHRG} shows the variation of the function  $f(\a, z)$ with the redshift
$z$ for several values of $\a$. We see that $\a $ should be at least 0.8, in order 
to allow for these OHRG to exist.  

\section{Summary and a combined constraint}

The main results of the present paper along the with constraints obtained
from the SNe Ia data \cite{abha} are summarized in Table 2. The motivation
for our work was to establish the viability of a linear coasting cosmology $a(t) = t$.    
Using gravitational lensing statistics, we find that
such a coasting is accommodated  within $1 \sigma$.  
The age determination of OHRG gives as a lower bound $\a \ge 0.8$ for a power law cosmology
$a(t) \propto t^{\a}$. 
Moreover, for such a cosmology $ H_0 t_0 = \a$.  
With updated value of $H_0 = 70 \pm 7$ km/sec/Mpc and $t_0 = 14 \pm 2$ Gyr \cite{pont},
it gives $\a = 0.98 \pm 0.25$. Dev et al (2001) reported that $\a = 1.0$ is consistent
with SNe Ia data (within 68\% confidence level).         
We find that $\a = 1.0$ is in concordance with the listed
observational tests. It is interesting to observe that the lensing
analysis barely accomodates an Einstein-de Sitter universe ($ \a = 2/3$)
at $2\sigma$. Similarly, the age determination of OHRG  also rules out $\a = 2/3$. We
conclude that the coasting cosmology with 
strictly linear evolution of scale factor, $a(t) = t$, cannot be ruled
out on the basis of these observations.    

\section*{Acknowledgments}

The authors are grateful to T.~D. Saini, A. Habib and N. Mahajan 
for useful discussions during the course of this work. We also thank
the referee for the useful comments.

\begin{table}
\begin{center}
\begin{tabular}{|c|ccccc|}\hline\hline
$Survey$ & $ \tilde \alpha$ & $ \gamma$ & $ v^{*} (Km /s)$ & $\phi^{*}
(Mpc^{-3})$ & $ F^{*}$
\\ \hline
\hline
$LPEM$ & $+ 0.2$ & $4.0$ &$205.3$ &$3.2\pm 0.17h^3 \times 10^{-3}$& $0.010$
\\ \hline
\end{tabular}
\caption{Lens and Schechter parameters for E/S0 galaxies.}
\label{T1}
\end{center}
\end{table}

\begin{table}
\begin{center}
\begin{tabular}{|l|l|r|}\hline\hline
Method &  Reference & $\a$  \\ 
\hline
\hline
Lensing Statistics &  & \\
& & \\
(i) $n_{\rm L}$ & This paper &  $1.09 \pm 0.3$ \\
(ii) Likelihood & This paper & $1.13^{+0.4}_{-0.3}$\\
\hspace{0.5cm} Analysis  &  &   \\
& & \\
OHRG & This paper & $\ge 0.8$ \\
& & \\
SNe Ia & Dev et al.(2001)& $1.004\pm0.043$ \\
\hline 
\end{tabular}
\caption{Constraints on $\a$ from various cosmological tests.}
\label{T2}
\end{center}
\end{table}
 
\begin{thebibliography}{99}
\bibitem{alternative} 
Carvalho, J.~C., Lima, J.~A.~S., \& Waga, I. 1992, \prd {\bf 46},
2404 ; Kraus, L.~M., \&  Turner, M. 1995, \grg, {\bf 27}, 1137 ; 
Lima, J.~A.~S., Germano, A.~S., \&  Abramo, L.~R.~W. 1996, \prd, {\bf 5},
4287;  Caldwell, R.~R., Dave, R., \& Steinhardt, P.~J. 1998, \prl, {\bf 80},
1582 .

\bibitem{Krauss} Krauss L. 1997, \apj, {\bf480}, 486.

\bibitem{models} Weinberg, S. 1989, {\it Rev. Mod. Phys.}, {\bf 61} ; 
Dolgov, A.~D., in the {\it The Very Early Universe}, eds. Gibbons, G.,
Siklos, S., Hawking, S.~W.,  C.~U. Press, 1982; Dolgov, A.~D. 1997, \prd,
{\bf 55}, 5881 (1997); Ford, L.~H. 1987, \prd, {\bf 35}, 2339.

\bibitem{Maoz1} Maoz, D.,  \& Rix, H.~W. 1993, \apj, {\bf 416}, 425.
 
\bibitem{Cooray} Cooray, A.~R., Quashnock, J.~M.,\&  Miller, M.~C. 1999,
\apj, {\bf 511}, 562.

\bibitem{Kolb} Kolb, E.~W. 1989, \apj, {\bf 344}, 543. 

\bibitem{allen} Allen, R.~E., Preprint No. {\bf astro-ph/9902042}.

\bibitem{mann} Manheim, P., \& Kazanas, D. 1990, \grg, {\bf 22}, 289.

\bibitem{meetu}
Lohiya, D., \& Sethi, M., 1999, {\it Class. Quan. Grav.}, {\bf 16}, 1545.

\bibitem{perl} Perlmutter, S., et al., {\bf astro-ph/9812133}.

\bibitem{abha} Dev, A., Sethi, M., \& Lohiya, D.  2001, {\it Phys. Letters B},
{\bf 504}, 207 .

\bibitem{annu} Batra, A., Lohiya, D., Mahajan, S., \&  Mukherjee, A. 2000, {\it Int. J.
Mod. Phys.}, {\bf D9}, 757; Batra, A., Sethi, M., \& Lohiya, D. 1999, \prd, {\bf 60},
108301. 

\bibitem{HST} Maoz, D., et al. 1993, \apj, {\bf 409}, 28 (Snapshot).

\bibitem{Crampton} Crampton, D., McClure, R.~D., \& Fletcher, J.~M. 1992,
\apj, {\bf 392}, 23.

\bibitem{Yee} Yee, H.~K.~C., Filippenko, A.~V., \& Tang, D. 1993, \aj, {\bf
105}, 7.

\bibitem{Surdej} Surdej, J., et al. 1993, \aj, {\bf 105}, 2064.

\bibitem{Jaunsen} Jaunsen, A.~O., Jablonski, M., Petterson, B.~R., \&
Stabell, R. 1995, \AnA, {\bf 300}, 323.

\bibitem{FKS} Kochanek, C.~S., Falco, E.~E., \& Schild, R. 1995, \apj, {\bf
452}, 109.

\bibitem{Bahcall} Bahcall, J.~N., et al. 1992, \apj,{\bf 387}, 56.

\bibitem{Maoz} Maoz, D., et al. 1993a, \apj, {\bf 402}, 69.

\bibitem{Maoz2} Maoz, D., et al. 1993b, \apj, {\bf 409}, 28.

\bibitem{TOG} Turner, E.~L., Ostriker, J.~P., \& Gott III, J.~R. 1984, \apj,
{\bf 284}, 1.

\bibitem{ffkt} Fukugita, M., Futamase, T., Kasai,
M., \& Turner, E.~L. 1992, \apj,  {\bf 393}, 3

\bibitem{loveday} Loveday, J.,Peterson, B.~A., Efstathiou, G.,\& Maddox,
S.~J. 1992, \apj,  {\bf 390}, 338 (LPEM).

\bibitem{c} Chiba, M. \& Yoshii, Y. 1999, \apj, {\bf 510}, 42.

\bibitem{lc} Christlein, D. 2000, \apj, {\bf 545}, 145.

\bibitem{blanton} Blanton, M. R., et al. 2001, \apj, {\bf 121}, 2358.

\bibitem{cross} Cross, N., et al. 2001, \mnras, {bf 324}, 825.

\bibitem{Turner} Turner, E.~L. 1990, \apjl, {\bf 365}, L43; Fukugita, M., \&
Turner, E.~L. 1991, \mnras, {\bf 253}, 99; Fukugita, M., Futamase, T., Kasai,
M., \& Turner, E.~L. 1992, \apj,  {\bf 393}, 3.

\bibitem{1CSK} Kochanek, C.~S. 1993, \apj, {\bf 419}, 12 [{\bf K93}].

\bibitem{2CSK} Kochanek, C.~S. 1996, \apj, {\bf 466}, 638 [{\bf K96}].

\bibitem{Qdata} Hartwich, F.~D.~A., \& Schade, D. 1990, {\it Ann. Rev.
Astron. Astrophys.},  {\bf 28}, 437.

\bibitem{Lima} Alcaniz, J.~S., \& Lima, J.~A.~S. 1999, \apj, {\bf 521}, L87.

\bibitem{Lima2} Lima, J.~M.~S., \& Alcaniz, J.~S. 2000, \mnras, {\bf 000},1. 

\bibitem{Dunlop} Dunlop, J. 1996, \nature, {\bf 381}, 581.

\bibitem{RH} Roos, M. and Harun-or-Raschid, S.~M., Preprint No. {\bf astro -
ph/0003040} (2000) .

\bibitem{Stockton} Stockton, A., Kellogg, M. \& Rirgway, S.~E. 1995, \apj,
{\bf 443}, L69.

\bibitem{Spinrard} Spinrard, S., et al. 1997, \apj, {\bf 484}, 581. 

\bibitem{Dunlop2} Dunlop, J. in {\it The Most Distant Radio Galaxies},
ed. H.~J.~A. Rottgering, P. Best \& M.~D. Lehnert, Dordrecht: Kluwer,
71 (1999). 

%\bibitem{wendy} Freedman, W.~L., Mould, J.~R., Kennicutt, R.~C., \&
%Madore, B.~F., {\bf astro-ph /9801080}.

\bibitem{Freedman} Freedman, W.~L. 2000, {\it Phys. Rep.}, {\bf 333}, 13.

\bibitem{Riess} Riess, A., et al. 1998, \aj, {\bf 116}, 1009.

\bibitem{pont} Pont, F., Mayor, M., Turon, C., \& Vandenberg, D. A. 1998,
\AnA, {\bf 329}, 87.

\end{thebibliography}
\end{document}